\documentclass[11pt,a4paper]{emulateapj}
\usepackage{epsfig}

\def\green{f_{_{\rm G}}}
\def\Green{F_{_{\rm G}}}
\def\sigmaT{\sigma_{\rm T}}
\def\xmax{x_{\rm max}}
\def\xmin{x_{\rm min}}
\def\freq{\nu_f}

\def\sgreen{f_{_{\rm G}}^{\rm S}}

\begin{document}

\title{A Fourier Transformed Bremsstrahlung Flash Model for the \break Production of
X-Ray Time Lags in Accreting Black Hole Sources}
\author{
  \text{John J. Kroon$^1$ and Peter A. Becker$^1$}\\
  \text{$^1$School of Physics, Astronomy, and Computational Sciences, George Mason University,} \\
  \text{Fairfax, VA 22030-4444; jkroon@gmu.edu, pbecker@gmu.edu}
}
\date{\today}

%
%

\begin{abstract}
Accreting black hole sources show a wide variety of rapid time
variability, including the manifestation of time lags during X-ray
transients, in which a delay (phase shift) is observed between the
Fourier components of the hard and soft spectra. Despite a large body of
observational evidence for time lags, no fundamental physical
explanation for the origin of this phenomenon has been presented. We
develop a new theoretical model for the production of X-ray time lags
based on an exact analytical solution for the Fourier transform
describing the diffusion and Comptonization of seed photons propagating
through a spherical corona. The resulting Green's function can be
convolved with any source distribution to compute the associated Fourier
transform and time lags, hence allowing us to explore a wide variety of
injection scenarios. We show that thermal Comptonization is able to
self-consistently explain both the X-ray time lags and the steady-state
(quiescent) X-ray spectrum observed in the low-hard state of Cyg X-1.
The reprocessing of bremsstrahlung seed photons produces X-ray time lags
that diminish with increasing Fourier frequency, in agreement with the
observations for a wide range of sources.
\end{abstract}

\section{Introduction}

The high energy emission from many accretion-powered X-ray binaries is
characterized by strong variability on timescales extending from
milliseconds to weeks or longer. The softest X-ray emission (including a
very soft thermal component) is observed during phases of rapid
accretion, and a hard X-ray component (along with associated strong
radio emission) is observed during periods of relatively slow accretion.
The quiescent spectrum in the hard state is characterized by a power-law
X-ray continuum, which is most likely due to the thermal Comptonization
of soft X-rays in a hot ($\sim 10^{8-9}\,$K) corona (e.g., Shapiro,
Lightman, \& Eardley 1976; Sunyaev \& Titarchuk 1980).

The X-ray variability observed on millisecond timescales allows us to
glimpse the inner workings of the accretion disk, but at these short
timescales, the signal-to-noise ratio is too low to study the
variability using snapshots of the spectrum. A useful alternative is
provided by the study of X-ray time lags, which are computed by
combining the Fourier transformed data streams in two energy channels.
Fourier-based analysis is extremely efficient because all of the data in
the observational time window is utilized, significantly reducing the
uncertainty in the derived parameters.

The physical origin of the observed X-ray time lags is not yet
completely understood. The time lags may be the result of the
upscattering of soft seed photons by hot electrons, known as ``Compton
reverberation'' (Payne 1980), which naturally produces hard time lags
because the higher-energy photons spend more time upscattering in the
plasma before escaping. This concept was explored by van der Klis et al.
(1987) and Miyamoto et al. (1988), who concluded that the observed
decrease in the time lag with increasing Fourier frequency in sources
such as Cyg X-1 was inconsistent with the Comptonization model. However,
the idea was reexamined later by Nowak et al. (1999a) and Hua, Kazanas,
\& Cui (1999, hereafter HKC), who used more sophisticated simulations
that produced better agreement with the data, but the size of the
scattering cloud and the implied energy budget remained problematic.

In this Letter, we reconsider the possible role of thermal
Comptonization in the formation of the X-ray time lags observed from
black hole candidates. We develop a rigorous method for the computation
of X-ray time lags based on the exact solution for the Fourier transform
of the Green's function describing the radiation signal emerging from a
spherical Compton scattering corona. By exploiting the linearity of the
problem, we are able to explore a wide variety of injection scenarios,
providing a useful alternative to Monte Carlo simulations, which are
often either unreliable or noisy at high Fourier frequencies (and short
timescales) due to sampling errors. We confirm that the Comptonization
of monochromatic seed photons injected into a homogeneous scattering
cloud is unable to account for the observed dependence of the X-ray time
lags on the Fourier frequency. However, we find that the reprocessing of
{\it broadband} (bremsstrahlung) injection reproduces the observed
frequency dependence within the constraints of an acceptable cloud size
and energy budget.

\section{Fourier-Based Time Lags}

In the complex cross-spectrum (CCS) method first introduced by van der
Klis et al. (1987), intensity variations in two energy channels
$\epsilon_{\rm hard}$ and $\epsilon_{\rm soft}$ are used to extract
phase shifts and associated time lags. The CCS method is implemented by
Fourier transforming the light curves in the hard and soft energy
channels, $h(t)$ and $s(t)$, respectively, to obtain the corresponding
Fourier transformed light curves $H(\freq)$ and $S(\freq)$, where
$\freq$ is the Fourier frequency. The two transformed light curves are
used to construct the complex cross spectrum, defined by
$C=S^{*}(\freq)H(\freq)$, and the resulting time lag is given by
\begin{equation} 
\delta t=\frac{\arg(C)}{2\pi\freq}=\frac{\arg \left(S^{*}H\right)}
{2 \pi \freq}
\ ,
\label{eq1}
\end{equation}
where $\arg(C)$ gives the phase lag (Nowak et al. 1999a). One can easily
demonstrate that if $s(t)$ and $h(t)$ represent the signal levels in the
soft and hard channels, respectively, and if a perfect time delay
between the two channels is introduced, such that $h(t)=s(t-\Delta t)$,
where $\Delta t > 0$, then the time lag derived using the CCS method
described above is $\delta t =\Delta t$, as expected. It is important to
note that time lags are only created during transients, and are never
produced during the steady-state reprocessing of radiation in a
scattering cloud with constant properties.

\section{Radiative Transfer Model}

Our focus here is on the determination of the effect of electron
scattering on the signal measured by a distant observer during a
transient in which a sudden flash of seed radiation is injected into a
hot corona overlying an accretion disk. In this first exact treatment of
the process, we restrict our attention to the case of a spherical,
homogeneous, and isothermal cloud. Once the solution for the Fourier
transform of the Green's function is in hand, we can compute the
corresponding theoretical prediction for the X-ray time lags using
(\ref{eq1}).

The reprocessing of $N_0$ seed photons injected simultaneously at
radius $r_0$ and time $t_0$ with energy $\epsilon_0$ is governed
by the time-dependent transport equation (e.g., Becker 1992, 2003)
\begin{eqnarray}
{\partial\green\over\partial t} &=&
{n_e \sigmaT c \over m_e c^2}
{1 \over \epsilon^2}{\partial\over\partial\epsilon}\big[\epsilon^4
\big(\green + kT_e{\partial \green \over\partial\epsilon}\big)\big]
+ {1 \over r^2} {\partial\over\partial r}\big(\kappa_0 r^2
{\partial \green \over\partial r}\big)
\nonumber
\\
&+&{N_0 \, \delta(\epsilon-\epsilon_0)\delta(r-r_0)\delta(t-t_0)
\over 4\pi r_0^2\epsilon_0^2}
\ ,
\label{eq2}
\end{eqnarray}
where $\green(\epsilon,r)$ is the Green's function describing the
distribution of photons with energy $\epsilon$ at radius $r$ inside the
cloud, and $T_e$ and $n_e$ denote the electron temperature and number
density, respectively. The terms on the right-hand side of (\ref{eq2})
represent thermal Comptonization, spatial diffusion, and photon sources,
respectively, and $\kappa_0=c/(3 n_e \sigmaT)$ denotes the spatial
diffusion coefficient. The Green's function is related to the total
photon number density, $n_r$, via
\begin{equation}
n_r(r) = \int_0^\infty \epsilon^2 \, \green(\epsilon,r) \, d\epsilon
\ .
\label{eq3}
\end{equation}

In the case of a homogeneous, isothermal scattering corona considered
here, $\kappa_0$ and $T_e$ are both constants, and it is convenient to
work in terms of the dimensionless energy $x \equiv \epsilon / kT_e$,
the dimensionless temperature $\Theta \equiv k T_e / m_e c^2$, the
dimensionless time $p \equiv t (c^2/3\kappa_0)$, and the scattering
optical depth $\tau \equiv n_e \sigmaT r=(c/3\kappa_0)r$, in which case
the transport equation~(\ref{eq2}) can be rewritten as
\begin{eqnarray}
{\partial\green\over\partial p} &=& {1 \over 3\tau^2} {\partial\over \partial \tau}
\big(\tau^2 {\partial \green \over \partial \tau}\big)
+ {\Theta \over x^2}{\partial \over \partial x}
\big[x^4\big(\green + {\partial \green \over \partial x}\big)\big]
\nonumber
\\
&+& {N_0\,\delta(x-x_0)\delta(\tau-\tau_0)\delta(p-p_0) \over 4\pi \tau_0^2
x_0^2 (m_e c^2)^3 \Theta^3 \ell_0^3}
\ ,
\label{eq4}
\end{eqnarray}
where $\tau_0$ is the injection optical depth, $x_0$ is the
dimensionless injection energy, $p_0$ is the dimensionless injection
time, and $\ell_0=3 \kappa_0 / c$ denotes the (constant) electron
scattering mean free path in the corona.

\section{Fourier Analysis}

We define the Fourier transform pair
$(\green,\Green)$ using
\begin{eqnarray}
\Green(x,\tau,\omega) &\equiv& \int_{-\infty}^\infty e^{i \omega p}
\green(x,\tau,p) \, dp
\ ,
\nonumber
\\
\green(x,\tau,p) &\equiv& {1 \over 2 \pi} \int_{-\infty}^\infty e^{-i \omega p}
\Green(x,\tau,\omega) \, d\omega
\ ,
\label{eq5}
\end{eqnarray}
where $\omega=\freq(2 \pi \ell_0/c)$ is the dimensionless Fourier
frequency. We can operate on (\ref{eq4}) with $\int_{-\infty}^\infty
e^{i\omega p}dp$ to obtain
\begin{eqnarray}
- i \omega \Green &=& {1 \over 3\tau^2} {\partial\over \partial \tau}
\big(\tau^2 {\partial \Green \over \partial \tau}\big)
+ {\Theta \over x^2}{\partial \over \partial x}
\big[x^4\big(\Green + {\partial \Green \over \partial x}\big)\big]
\nonumber
\\
&+& {N_0\,\delta(x-x_0)\delta(\tau-\tau_0)e^{i \omega p_0} \over 4\pi \tau_0^2
x_0^2 (m_e c^2)^3 \Theta^3 \ell_0^3}
\ ,
\label{eq6}
\end{eqnarray}
where $i^2=-1$.

When $x \ne x_0$, (\ref{eq6}) is separable using $F_\lambda =
G(\lambda,\tau) H(\lambda,x)$ and this yields the differential equations
\begin{equation}
{1 \over \tau^2} {d \over d\tau}\big(\tau^2 {dG \over d\tau}\big)
+ \lambda G = 0
\ ,
\label{eq7}
\end{equation}
\begin{equation}
{1 \over x^2} {d \over dx}\big[x^4\big(H + {dH \over dx}\big)\big]
- {s \over 3 \Theta} H = 0
\ ,
\label{eq8}
\end{equation}
where $\lambda$ is the separation constant and $s \equiv \lambda - 3 i
\omega$. The solutions for $G$ and $H$ satisfying suitable boundary
conditions in energy and radius are
\begin{equation}
G(\lambda,\tau)={\sin(\tau\sqrt{\lambda}) \over \tau}
\ ,
\label{eq9}
\end{equation}
and
\begin{equation}
H(\lambda,x)=(x_0 x)^{-2} e^{-(x+x_0)/2} \, M_{2,\mu}(\xmin) \,
W_{2,\mu}(\xmax)
\ ,
\label{eq10}
\end{equation}
where $M_{2,\mu}$ and $W_{2,\mu}$ denote Whittaker functions,
the constant $\mu$ is given by
\begin{equation}
\mu \equiv \Big({9 \over 4} + {\lambda \over 3 \Theta} - {i \omega \over \Theta}
\Big)^{1/2}
\ ,
\label{eq11}
\end{equation}
and we have made the definitions $\xmin \equiv \min(x,x_0)$,
$\xmax \equiv \max(x,x_0)$.

The spherical scattering corona has a finite size, with outer radius
$r=R$. At the surface of the cloud, the distribution function $f$ must
satisfy the free-streaming boundary condition $-\kappa_0 (\partial f/
\partial r)=c\,f$. This expression yields a constraint on the spatial
separation function $G$, which can be written in terms of the optical
depth $\tau$ as
\begin{equation}
{1 \over 3}{dG \over d\tau} + G
\bigg|_{\tau=\tau_*} = 0
\ ,
\label{eq12}
\end{equation}
where $\tau_* \equiv n_e \sigmaT R = R/\ell_0$ is the scattering optical
thickness from the center to the edge of the corona. The roots of
(\ref{eq12}) are the eigenvalues $\lambda_n$. Since the eigenvalue
equation is transcendental, the eigenvalues must be determined using a
numerical root-finding procedure. The eigenvalues $\lambda_n$ are all
real and positive, and the corresponding values of $\mu$ are computed
using (\ref{eq11}) by setting $\lambda=\lambda_n$. The associated
eigenfunctions $G_n$ and $H_n$ are defined by
\begin{equation}
G_n(\tau) \equiv G(\lambda_n,\tau) \ , \ \ \ 
H_n(x) \equiv H(\lambda_n,x)
\ .
\label{eq13}
\end{equation}
Application of the Sturm-Liouville theorem shows that the spatial
eigenfunctions $G_n$ form a complete orthogonal set.

The exact solution for the Fourier transform $\Green$ of the Green's
function describing the time-dependent photon distribution can now be
expressed using the series
\begin{equation}
\Green(x,\tau,\omega)=\sum_{n=0}^\infty a_n G_n(\tau) H_n(x)
\ ,
\label{eq14}
\end{equation}
where the expansion coefficients $a_n$ can be calculated by exploiting
the orthogonality of the spatial eigenfunctions $G_n$, combined with the
derivative jump condition
\begin{equation}
\Delta\Big[{\partial \Green \over \partial x}\Big]
= {-N_0 \, \delta(\tau-\tau_0) e^{i \omega p_0} \over 4 \pi \tau_0^2 x_0^4
(m_e c^2)^3 \ell_0^3 \Theta^4}
\ ,
\label{eq15}
\end{equation}
obtained by integrating the transport equation~(\ref{eq6}) over a small
energy range around $x=x_0$. Solving for the expansion coefficients and
combining the result with (\ref{eq14}), we find that the exact
solution for the Fourier transform of the Green's function of the
time-dependent spectrum is given by
\begin{equation}
\Green = {N_0 \, e^{i \omega p_0} \, e^{x_0} \over 4 \pi \ell_0^3
(m_e c^2)^3} \sum_{n=0}^\infty {\Gamma(\mu\!-\!3/2) \, G_n(\tau_0)
G_n(\tau) H_n(x) \over  \Theta^4 \, \Gamma(1+2\mu) \, \mathcal{I}_n}
\ ,
\label{eq16}
\end{equation}
where the quadratic normalization integrals of the spatial
eigenfunctions are defined by $\mathcal{I}_n\equiv\int_0^{\tau_*} \tau^2
G^2_n(\tau)d\tau$, and the values of $\mu$ are computed by substituting
the eigenvalues $\lambda_n$ into (\ref{eq11}).

\section{Quiescent Spectrum}

In the integrated model considered here, both the steady-state
(quiescent) and the transient X-ray spectral components are produced via
thermal Comptonization in the corona, but the source of seed photons is
different in the two cases. We assume here that the steady-state
quiescent spectrum is the result of the upscattering in the corona of
soft seed photons continually injected into the corona from the
underlying cool disk (note however that reprocessed bremsstrahlung
emission produced in the corona itself may also contribute significantly
to the high-energy X-ray spectrum). We assume that the transient
component (responsible for the time lags) is the result of a sudden
flash of seed photons injected into the corona at a particular radius,
due to some instability, which may produce either monochromatic or
broadband emission.

We can tie down the fundamental cloud parameters $\Theta$ and $\tau_*$
by comparing the theoretical quiescent spectrum with the observational
X-ray data. The quiescent spectrum, $\sgreen(\epsilon,r)$, is computed
by solving the steady-state transport equation
\begin{eqnarray}
0 &=&
{n_e \sigmaT c \over m_e c^2}
{1 \over \epsilon^2}{\partial\over\partial\epsilon}\big[\epsilon^4
\big(\sgreen + kT_e{\partial \sgreen \over\partial\epsilon}\big)\big]
\nonumber
\\
&+& {1 \over r^2} {\partial\over\partial r}\big(\kappa_0 r^2
{\partial \sgreen \over\partial r}\big)
+ {\dot N_0 \, \delta(\epsilon-\epsilon_0)
\over (4/3) \pi R^3 \epsilon_0^2}
\ ,
\label{eq17}
\end{eqnarray}
where $\dot N_0$ denotes the rate at which photons with energy
$\epsilon_0$ are injected uniformly throughout the cloud of radius $R$.
Adopting the same dimensionless variables $x$ and $\tau$ used in
(\ref{eq4}) now yields
\begin{eqnarray}
0 &=& {1 \over 3\tau^2} {\partial\over \partial \tau}
\big(\tau^2 {\partial \sgreen \over \partial \tau}\big)
+ {\Theta \over x^2}{\partial \over \partial x}
\big[x^4\big(\sgreen + {\partial \sgreen \over \partial x}\big)\big]
\nonumber
\\
&+& {\dot N_0\,\ell_0\,\delta(x-x_0) \over (4/3)
\pi R^3 c (m_e c^2)^3 \Theta^3 x_0^2}
\ .
\label{eq18}
\end{eqnarray}
In analogy with
(\ref{eq6}), we can separate (\ref{eq18}) for $x \ne x_0$ in terms of
the functions $f_\lambda = G(\lambda,\tau) K(\lambda,x)$, where $G$ and
$K$ satisfy the differential equations
\begin{equation}
{1 \over \tau^2} {d \over d\tau}\big(\tau^2 {dG \over d\tau}\big)
+ \lambda G = 0
\ ,
\label{eq19}
\end{equation}
\begin{equation}
{1 \over x^2} {d \over dx}\big[x^4\big(K + {dK \over dx}\big)\big]
- {\lambda \over 3 \Theta} K = 0
\ .
\label{eq20}
\end{equation}
The solutions for $G$ and $K$ are given by
\begin{equation}
G(\lambda,\tau)={\sin(\tau\sqrt{\lambda}) \over \tau}
\ ,
\label{eq21}
\end{equation}
and
\begin{equation}
K(\lambda,x)=(x_0 x)^{-2} e^{-(x+x_0)/2} \, M_{2,\sigma}(\xmin) \,
W_{2,\sigma}(\xmax)
\ ,
\label{eq22}
\end{equation}
respectively, where
\begin{equation}
\sigma \equiv \Big({9 \over 4} + {\lambda \over 3 \Theta}\Big)^{1/2}
\ .
\label{eq23}
\end{equation}
Since (\ref{eq7}) and (\ref{eq19}) are identical, it follows that the
eigenvalues $\lambda_n$ are exactly the same ones obtained in the
solution for the Fourier transform $\Green$ in \S~4. The corresponding
eigenfunctions $G_n$ and $K_n$ are given by
\begin{equation}
G_n(\tau) \equiv G(\lambda_n,\tau) \ , \ \ \ 
K_n(x) \equiv K(\lambda_n,x)
\ .
\label{eq24}
\end{equation}

The exact solution for the steady-state photon distribution function
$\sgreen$ can now be expressed using the series
\begin{equation}
\sgreen(x,\tau)=\sum_{n=0}^\infty b_n G_n(\tau) K_n(x)
\ ,
\label{eq25}
\end{equation}
where the expansion coefficients $b_n$ are calculated using the same
approach used to obtain the coefficients $a_n$ in the expansion for
$\Green$. In this case, integration of (\ref{eq18}) yields the derivative
jump condition
\begin{equation}
\Delta\Big[{\partial \sgreen \over \partial x}\Big]
= {- \dot N_0 \, \ell_0 \over (4/3) \pi R^3 c (m_e c^2)^3 \Theta^4 x_0^4}
\ ,
\label{eq26}
\end{equation}
After some algebra, we find that the exact solution for the steady-state
(quiescent) Comptonized spectrum is given by
\begin{equation}
\sgreen(x,\tau) = {9 \dot N_0 \, e^{x_0} \over
4 \pi R^2}
\sum_{n=0}^\infty {\Gamma(\sigma\!-\!3/2) \, \sin(\tau_*\sqrt{\lambda_n})
G_n(\tau) K_n(x) \over \Theta^4 \, c \, (m_e c^2)^3 \lambda_n \ \Gamma(1+2\sigma)
\, \mathcal{I}_n}
\ ,
\label{eq27}
\end{equation}
where the values of $\sigma$ are computed by substituting
the eigenvalues $\lambda_n$ into (\ref{eq23}). The photon number
flux measured at the detector, ${\cal F}_\epsilon$, can be computed
from $\sgreen(x,\tau)$ using
\begin{equation}
{\cal F}_\epsilon(\epsilon) = \left(R \over D\right)^2 c \ \epsilon^2
\sgreen\left({\epsilon\over k T_e},\tau_*\right)
\ ,
\label{eq28}
\end{equation}
where $D$ is the distance to the source. The emergent spectrum will
be compared with the quiescent data for Cyg X-1 in \S~6.

\section{Application to Cyg X-1}

The integrated model we have developed can be applied to any active
galactic nucleus or accreting galactic black-hole candidate using a
two-step approach. First, the cloud temperature $\Theta$ and optical
thickness $\tau_*$ are determined by comparing the photon flux computed
using (\ref{eq28}) with the observed quiescent X-ray spectrum. Second,
the Fourier transform of the emergent signal is computed, and the
results are substituted into the CCS method using (\ref{eq1}) to
calculate the associated time lags.

As an example, we use the integrated model to interpret the X-ray
spectral and timing observations of Cyg X-1. In Figure~1a, the
steady-state spectrum (\ref{eq28}) is compared with the quiescent X-ray
spectrum observed in the low hard state of Cyg X-1 by Cadolle Bel et al.
(2006) assuming $D = 2.4\,$kpc. The solid curve in Figure~1a represents
the steady-state spectrum resulting from the thermal Comptonization of
monochromatic seed photons continually injected throughout the entire
corona with energy $\epsilon_0=0.1\,$keV, which approximates a
$T=10^6\,$K blackbody source. The temperature and optical thickness of
the corona are $\Theta=0.12$ and $\tau_*=2.5$, respectively.

In computing the associated time lags, we consider both monochromatic
injection and bremsstrahlung injection. The Fourier transform in the
bremsstrahlung case is computed using the integral convolution
\begin{equation}
F_{\rm brem}(x) = \int_{x_{\rm abs}}^\infty
\Green Q N_0^{-1} dx_0
\ ,
\label{eq29}
\end{equation}
where the bremsstrahlung source term is given by $Q \propto
e^{-x_0}/x_0$ and $x_{\rm abs}=\epsilon_{\rm abs}/(k T_e)$ is the
low-energy cutoff due to self-absorption in the source (Rybicki \&
Lightman 1979). The integral convolution in (\ref{eq29}) can be carried
out analytically but we do not include the result here for brevity. The
Fourier transform of the signal is computed using either (\ref{eq16}) or
(\ref{eq29}), for monochromatic or bremsstrahlung injection,
respectively, and then the time lags are calculated using (\ref{eq1}).

In Figure~1b, the theoretical time lags are plotted as a function of the
Fourier frequency $\nu_f$ and compared with the RXTE data obtained in
the hard state of Cyg X-1 by Nowak et al. (1999a). The time lags are
computed using the same values $\Theta=0.12$ and $\tau_*=2.5$ used to
calculate the quiescent spectrum in Figure~1a. The red dashed curve
corresponds to the injection of {\it monochromatic} seed photons with
energy $\epsilon_0=0.1$\,keV at optical depth $\tau_0=2.5$ into a cloud
with radius $R=10,000$\,km. The blue dot-dashed curve denotes the lags
resulting from {\it broadband} (bremsstrahlung) injection at optical
depth $\tau_0=2.5$ with $\epsilon_{\rm abs}=1.6$\,keV into a cloud with
radius $R=40,000$\,km. The black solid curve represents the lags
resulting from bremsstrahlung injection at optical depth $\tau_0=2.5$
with $\epsilon_{\rm abs}=1.6$\,keV into a cloud with radius
$R=30,000$\,km. The bremsstrahlung time lag results fit the
observational data fairly well, which may imply that the X-ray time lags
are produced at much larger radii than the broad iron lines.

\begin{figure}
\begin{center}$
\begin{array}{c}
\includegraphics[width=2.6in]{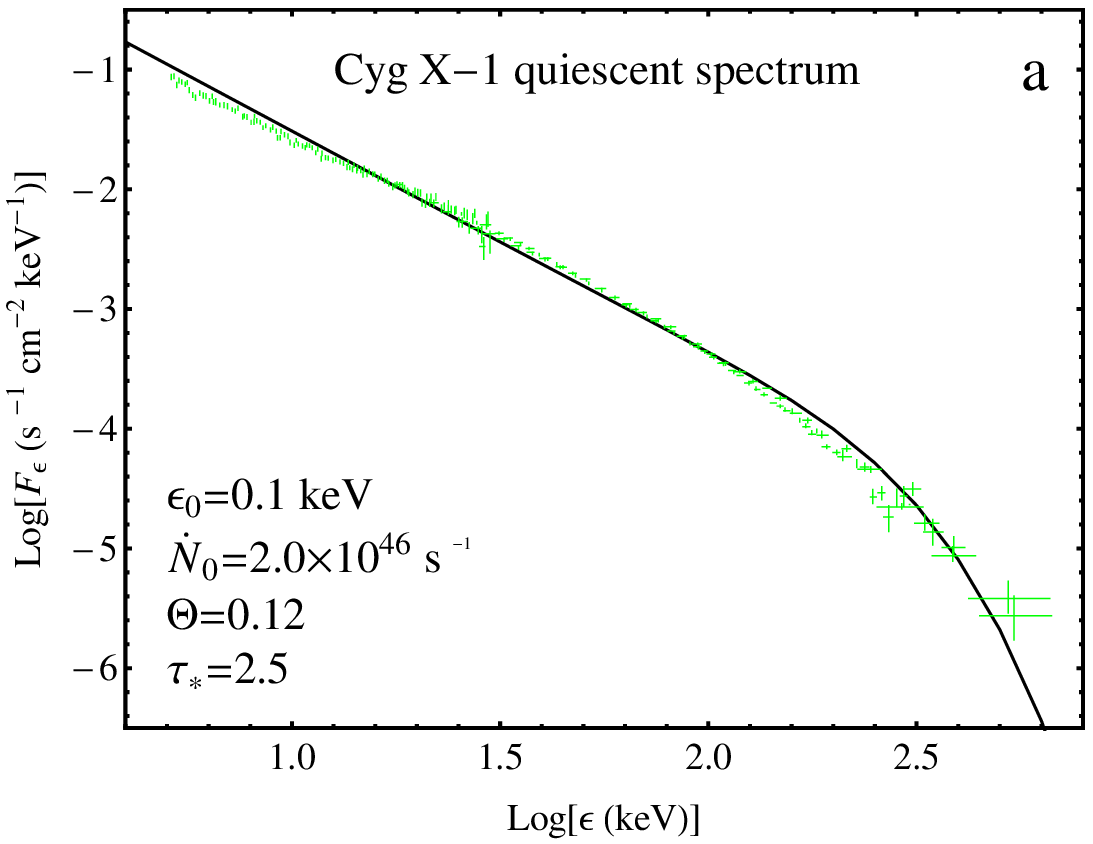}\\
\includegraphics[width=2.7in]{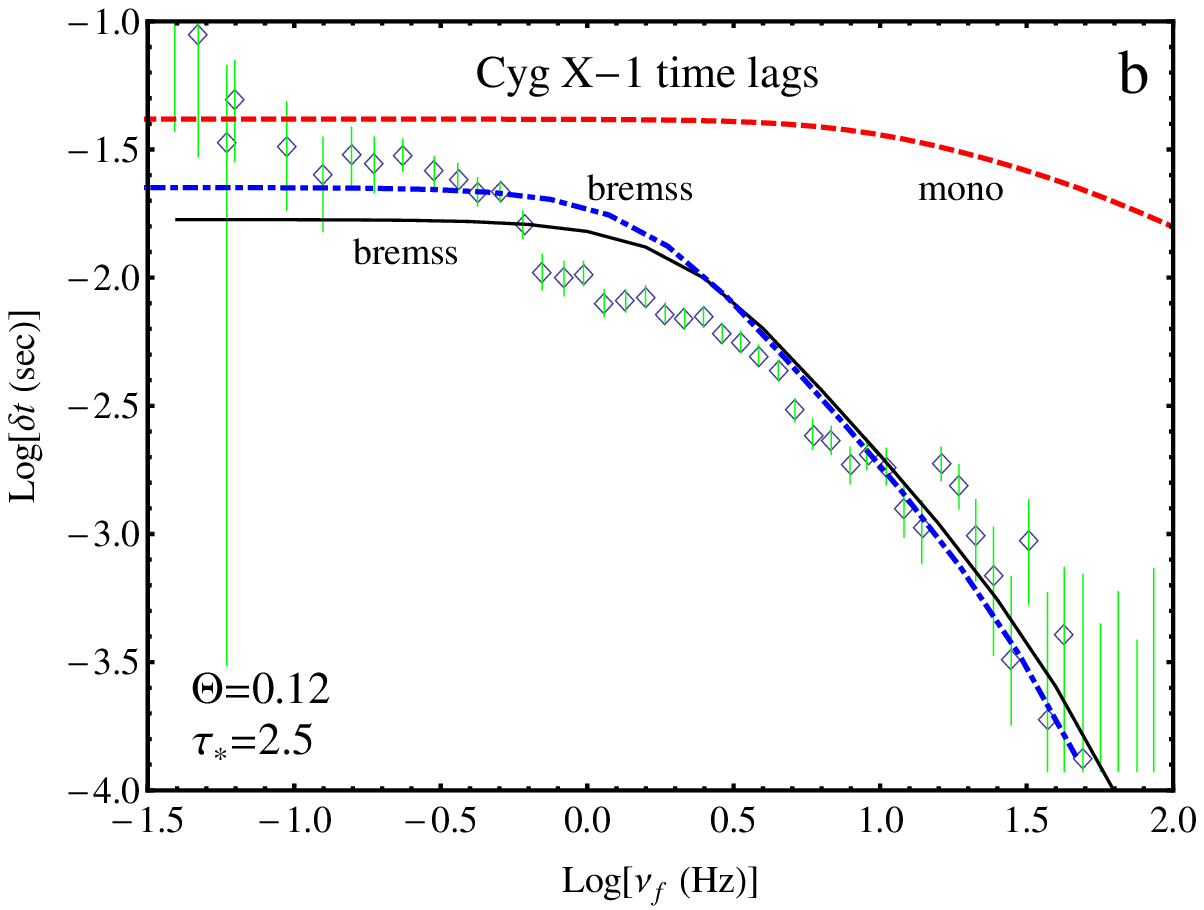}
\end{array}$
\end{center}
\label{fig:plots}
\caption{(a) Solid line represents the quiescent photon flux at the
detector computed using (\ref{eq28}) with $\Theta=0.12$, $\tau_*=2.5$,
$\dot N_0 = 2.0 \times 10^{46}\,{\rm s}^{-1}$, $\epsilon_0=0.1\,$keV.
Crosses represent the data from the July-November 2003 INTEGRAL
observation of Cyg X-1 in the hard state reported by Cadolle Bel et al.
(2006). (b) Time lags computed using (\ref{eq1}) based on three
different injection scenarios are compared with the data from a
transient of Cyg X-1 observed in October 1996 as discussed by Nowak et
al. (1999a). The theory curves were generated using $\epsilon_{\rm
soft}=2\,$keV and $\epsilon_{\rm hard}=11\,$keV.}
\end{figure}

\section{Discussion and Conclusion}

Based on the results plotted in Figure~1b, we conclude that thermal
Comptonization of monochromatic radiation is unable to reproduce the
observed time lag profiles, in agreement with the conclusions reached by
Miyamoto et al. (1988), Nowak et al. (1999b), and HKC. On the other hand,
the two curves in Figure~1b corresponding to the reprocessing of {\it
bremsstrahlung} seed radiation are a near match to the observed profiles
at high frequencies, although the time lags plateau at low frequencies
due to the homogeneous density distribution assumed here. HKC were able
to remedy the problem at low frequencies by focusing on an inhomogeneous
corona with electron density $n_e \propto 1/r$, although this approach
leads to other problems as discussed below.

Compton upscattering in a hot corona produces time lags that increase
logarithmically with increasing hard channel energy, and the
approximately logarithmic dependence seen in the data provided part of
the observational motivation for the simulations carried out by HKC. Our
results confirm the logarithmic behavior, as expected, but there are
some interesting differences between our results and those obtained by
HKC. Our model requires a relatively high electron temperature, $T_e
\sim 10^8\,$K, at large radii, $r \sim 10^3\,GM/c^2$, in order to
explain the observed time lags, in agreement with other Compton
scattering models (HKC; Poutanen 2001). This temperature is higher than
some disk models predict at that distance, although we note that You et
al. (2012) have developed a fully relativistic two-temperature
disk-corona model that predicts electron temperatures $T_e \sim 10^8\,$K
on size scales comparable to the radius of the corona in our model. The
heating problem is more severe in the model of HKC, which requires a hot
corona extending out to $\sim 10^{4-5}GM/c^2$, and it is unclear whether
the dissipation mechanism proposed by You et al. (2012) can provide the
required energy at that distance.

The qualitatively different behavior observed in the bremsstrahlung
injection scenario stems from the fact that photons with energies inside
both energy channel windows already exist in the injected spectrum.
Hence, when the injection occurs close to the edge of the cloud,
``prompt'' photons inside the energy channel windows are able to escape
immediately. The prompt escape phase represents the fastest timescale
during the transient, and consequently this is the process that
contributes to the high-frequency part of the Fourier transform. The
resulting time lag diminishes with increasing Fourier frequency because
the high- and low-energy prompt photons escape at the same rate, with no
upscattering required, so there is no relative delay between the two
channel energies at the highest Fourier frequencies.

Conversely, at low frequencies, the process becomes dominated by the
longest timescale phenomenon, which is the exponential decay of the
photon number density due to photons that remain in the plasma for a
long time. This phase is dominated by the upscattering of very soft
bremsstrahlung seed photons, which take longer to upscatter to the
higher energy channel. Hence the low-frequency behavior of the
bremsstrahlung time lag profile is similar to what is observed in the
case of monochromatic injection.

We have developed a rigorous analytical model with very few free
parameters that is capable of simultaneously describing the quiescent
X-ray spectrum from Cyg X-1, as well as the transient time lags, based on
thermal Comptonization in a homogeneous corona. In future work, we plan
to extend the analytical model to treat an inhomogeneous corona, with
electron density $n_e \propto 1/r$. The homogeneous case treated here
underestimates the time lags at low frequencies, and we expect the
inhomogeneous density to result in a better fit across the entire time
lag profile since the distribution of scattering times will be more
accurately computed.

The authors are grateful to the anonymous referee for a number of
insightful comments and suggestions.

\end{document}